%% file: main.tex
\pdfoutput=1
\newif\ifFull
\Fullfalse
\ifFull
\documentclass[11pt]{article}
\usepackage[margin=1in]{geometry}
\else
\documentclass{llncs}
\fi
\usepackage{graphicx}
\usepackage{cite}
\usepackage{url}
\usepackage{amsmath,amsfonts,amssymb,latexsym}
\usepackage{parskip}
\usepackage[noend]{algorithmic}

\usepackage{multirow}

\begin{document}

\ifFull\else
\pagestyle{plain}
\renewcommand{\subsection}[1]{\paragraph{\textbf{#1}.}}
\fi

\title{J-Viz: Sibling-First Recursive Graph Drawing for 
 Visualizing Java Bytecode}

\author{Md.~Jawaherul Alam \and Michael T.~Goodrich \and Timothy Johnson}

\institute{
Dept. of Computer Science, University of California, Irvine, CA USA 
}

\date{}

\maketitle

\begin{abstract}
We describe a graph visualization tool for visualizing Java bytecode.
Our tool, which we call J-Viz, visualizes connected directed graphs
according to a canonical node ordering,
which we call the \emph{sibling-first recursive} (SFR) numbering.
The particular graphs we consider are derived from applying
Shiver's $k$-CFA framework to Java bytecode, and our visualizer includes
helpful links between the nodes of an input graph and the Java bytecode that
produced it, as well as a decompiled version of that Java bytecode.
We show through several case studies that
the canonical drawing paradigm used in
J-Viz is effective for identifying 
potential security vulnerabilities
and repeated use of the same code in Java applications.
\end{abstract}

\section{Introduction}
The Space/Time Analysis for Cybersecurity (STAC) program~\cite{stac}
at the U.S.~Defense Advanced Research Projects Agency (DARPA)
aims to develop new program analysis techniques and tools for 
identifying vulnerabilities related to the space and time 
resource usage behavior of algorithms, and
specifically to vulnerabilities based on algorithmic complexity and 
side channel attacks. 
STAC seeks to enable security 
analysts to identify algorithmic resource usage
vulnerabilities in software 
to support a methodical search for them in the software upon
which the U.S. government, military, and economy depend~\cite{stac}.

\subsection{Our Contributions}
In this paper, we describe a tool, 
the JVM abstractng abstract machine (Jaam) Visualizer, or
``J-Viz'' for short, which is intended for 
use by security analysts to perform such searches through the exploration
of graphs derived from Java bytecode.
We are not attempting to 
solve the problem of identifying software algorithmic complexity attacks
in a completely automated way, but instead we are providing a means
for doing semi-automated analysis that
increases the efficiency of a human analyst. 
The workflow for our tool involves taking
a given program, specified in Java bytecode, and constructing 
a control-flow graph of the possible execution paths for this software,
using a framework known as \emph{control flow analysis} (CFA)~\cite{Shivers-thesis}. 
Our tool then provides a human analyst with an
interactive view of this graph, including heuristics
for aiding the identification
of which parts of the provided program seem suspicious.

One of the main components of our J-Viz tool involves visualizing
control-flow graphs in a canonical way based on a novel vertex 
numbering scheme that we call the \emph{sibling-first recursive} numbering.
This numbering scheme is essentially a hybrid between the well-known 
breadth-first and depth-first numbering
schemes, but differs from both in a way that appears
 to be more useful for visualizing control-flow graphs.
 In particular, as we show in some case studies, this approach tends
 to highlight areas in software where code is repeated and it also allows us
 to provide visual highlights of code that is contained in deeply
 nested loops.
We designed J-Viz with the following goals in mind:

\begin{itemize}
\item
We want users to be easily able to recognize patterns in source
code from our visualizations. Thus, similar sections of code should produce
similar subgraphs, which should be drawn in a similar way.

\item
We want to use a hierarchical visualization, in which users can
collapse or expand sections of the graph to different levels of
detail. But we also want them to be able to build a consistent
mental model of the graph. Thus, drawings should not drastically
shift the relative positions of the vertices when sections are
collapsed or expanded.

\item No matter what sequence of actions the user performs, drawings
should be consistent. That is, the same view of a graph, in which
the same set of nodes are collapsed and expanded, should always be
drawn in the same way.

\item Our system should rank sections of the graph by how likely they are
to produce vulnerabilities, and display this information visually
to the user.
\end{itemize}
We believe that J-Viz makes substantial progress in achieving these goals,
and we provide several case studies in this paper that support this
conclusion.
 
 \subsection{Related Work}
 Although it is using different means to achieve mental map preservation,
 the J-Viz system follows in a long line of research
 on techniques directed at preserving the mental map of a graph drawn
 dynamically.
 For instance,
 Misue {\it et al.}~\cite{MISUE1995183} discuss node movement 
 adjustments, including avoiding node overlaps, 
 for preserving the mental map.
 Diehl and Carsten~\cite{Diehl2002} discuss force-directed approaches for
 preserving the mental map between instances of a changing graph.
 Goodrich and Pszona~\cite{Goodrich2013} study efficient algorithms for 
 minimizing vertex movements as a graph is incrementally revealed in
 an online manner.
 With respect to existing software systems,
 the Graphviz~\cite{Ellson2004}
 and GraphAEL~\cite{Erten2004}
 systems both include algorithms intended to preserve
 the mental map as a graph is modified.
 Bridgeman and Tamassia~\cite{Bridgeman1998} formally study metrics for
 characterizing mental map preservation between different instances of a
 changing graph.
 So as to provide an empirical basis for such work,
 a user study of Purchase {\it et al.}~\cite{Purchase2007} supports 
 the thesis that preserving the mental map for graph visualization is a useful
 goal to aid users in performing tasks on graphs.
 
Visualization tools have also previously been applied to source
code. Doxygen~\cite{doxygen}, a tool for automatically generating
documentation, can produce various kinds of diagrams for visualizing
code, including call graphs. It is generally configured to use the
\textit{dot}~\cite{dot} tool from GraphViz to draw these graphs
hierarchically. Similarly, Visual Studio can visualize call graphs
to aid programmers in debugging applications~\cite{code-maps}.
In constrast with these systems,
our J-Viz tool provides four main features that these tools do not provide. 
First, J-Viz shows a greater level of detail, since
it analyzes code at the level of individual instructions rather
than methods. Second, J-Viz allows the user to interact with a graph
and produce multiple views of the same Java bytecode. Third, 
the layout
algorithm used in J-Viz is designed to draw similar code fragments
in the same (canonical) way, so as to highlight portions of repeated code. 
Fourth, J-Viz guides the user to potential security vulnerabilities,
by highlighting nodes that are believed 
to be risky based on algorithmic complexity (or other factors), whereas
these other systems were not focused on software security.

Another tool,
Jinsight~\cite{jinsight}, can be used to profile
 a Java program to provide various
views of resource usage, such as highlighting which instances of a
class have taken the most time or used the most memory. This tool does
not provide a full graph of the program's possible execution paths,
however, which we believe to be essential for 
detecting security vulnerabilities.

\section{Graph Generation via Static Analysis}
In this section, we review the process that takes Java bytecode
as input and produces the graphs that are visualized in J-Viz.
These graphs are produced using the \emph{JVM abstracting abstract machine}
(\emph{Jaam}) tool~\cite{jaam}
developed at the University of Utah based on the work
of Van~Horn and Might~\cite{HM11}, which itself is 
based on control-flow analysis (CFA)
framework known as \emph{$k$-CFA}~\cite{Shivers-thesis,Matt-thesis}. 
Because there could be exponentially many possible execution paths of any
given program, which would be too large to visualize and reason about,
the $k$-CFA framework compresses execution paths into a graph of 
reasonable size that represents possible executions of a Java program
at the instruction level. Such a graph is called \emph{sound} if
it represents every possible execution path, and \textit{precise} if
it excludes every impossible execution path. 
The $k$-CFA framework is sound, and it has a tunable degree of precision
based on the integer parameter, $k$, albeit at the cost of
creating additional states in the graph for larger values of $k$.

At the lowest level of the hierarchy, 0-CFA, we discard contextual
information and generate one state, which forms a vertex in the graph
representation, for each line of Java bytecode. Then
we add edges for every possible state that could be reached from a
given state. For example, a return statement will have an edge to
every place from which our current method could have been called.
At the next level, 1-CFA, each state also tracks the location from
which its method was called. (See Fig.~\ref{fig:Fib-drawing} in Appendix~\ref{sec:examples}
for example graphs produced by 1-CFA.) This easily generalizes to higher levels, so
that for $k$-CFA, each state stores the locations of the previous
$k$ function calls. This added information allows many of the
spurious branches produced by 0-CFA to be pruned.
(For additional information, please see 
more detailed descriptions of $k$-CFA~\cite{HM11,Shivers-thesis,Matt-thesis}.)

0-CFA is known to take $O(n^3)$ time to construct a graph for a
program with $n$ lines of code, and this is believed to be
tight~\cite{cubic-bottleneck}.  $k$-CFA is EXPTIME-complete for
functional languages~\cite{VanHorn:2008}, but can be solved in
polynomial time for object-oriented languages~\cite{MSH10}. 
Thus, to provide 
a reasonable balance between soundness, precision, and efficiency,
the version of the Jaam static analyzer used for the work of this paper 
is based on 1-CFA.
To summarize, then,
the Jaam static analyzer 
takes as input Java bytecode for a given program and 
produces an directed graph, $G$, that represents the results
of a 1-CFA performed on this bytecode.
This graph is \emph{ordered}, in the sense that the outgoing edges for 
each node are sorted according to the order in which the 
corresponding instructions appear in the Java bytecode.

\section{Our Sibling-First Recursive Layout Algorithm}
\begin{figure}[hbt]
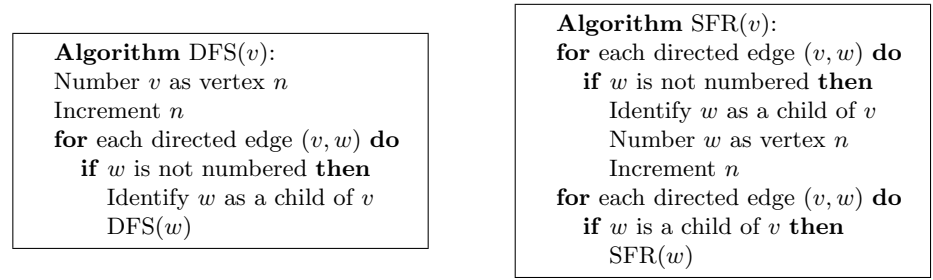

\begin{center}
\begin{tabular}{c@{\hspace*{.45in}}c}
\framebox{
\begin{minipage}{2in}
\begin{algorithmic}
\STATE \textbf{Algorithm} DFS$(v)$:
\STATE Number $v$ as vertex $n$ 
\STATE Increment $n$
\FOR {each directed edge $(v,w)$}
\IF {$w$ is not numbered}
\STATE Identify $w$ as a child of $v$
\STATE DFS$(w)$ 
\ENDIF
\ENDFOR
\end{algorithmic}
\end{minipage}
}
&
\framebox{
\begin{minipage}{2in}
\begin{algorithmic}
\STATE \textbf{Algorithm} SFR$(v)$:
\FOR {each directed edge $(v,w)$}
\IF {$w$ is not numbered}
\STATE Identify $w$ as a child of $v$
\STATE Number $w$ as vertex $n$ 
\STATE Increment $n$
\ENDIF
\ENDFOR
\FOR {each directed edge $(v,w)$}
\IF {$w$ is a child of $v$}
\STATE SFR$(w)$
\ENDIF
\ENDFOR
\end{algorithmic}
\end{minipage}
}
\end{tabular}
\end{center}
\caption{
\label{fig:dfs}
   DFS and SFR algorithms to explore the connected component of 
   a vertex, $v$, in
   a directed graph, $G$. We assume there is a global variable, $n$,
   which is used to number the vertices.
   In the case of DFS, we initialize $n=1$ and call DFS$(v)$ on a vertex $v$
   that is to become the root of the DFS tree.
   In the case of SFR, we initialize a vertex, $v$,
   as the root of the SFR tree, numbering it as vertex 1,
   and we set $n=2$ and call SFR$(v)$.}
\end{figure}

Our approach to the layout of graphs produced by the Jaam tool~\cite{jaam}
is based on what we believe is a novel graph numbering scheme, which we
call a \emph{sibling-first recursive} (SFR) numbering.
Intuitively, SFR is a hybrid numbering scheme that combines features of
a breadth-first search (BFS) numbering and a depth-first search (DFS) numbering.
We show in Figure~\ref{fig:dfs} the difference between 
algorithms for doing a depth-first search
(DFS) ordering of a directed graph
and a sibling-first recursive (SFR) numbering. 
See also Figures~\ref{fig:sfr-number},
\ref{fig:sfr}, and~\ref{fig:factorial-drawing} in the appendix,
which illustrate SFR spanning trees and their differences with DFS and BFS
spanning trees.

Our motivation for using the SFR numbering 
is that we feel it produces a rooted spanning tree that corresponds more
intuitively with the way that programmers conceptualize the main
``backbone'' of the control flow of their software.
For example, it places the true-false branches of if statements as children
of the condition that branches to them. 
In addition,
it places the multiple branches of a switch statement as children of the
condition that branches to them, even if some of the branches flow-through to
other branches.
(E.g., see Fig.~\ref{fig:layouts}.)
Furthermore SFR numbering also enables the viewer to visually identify
 isomorphic subgraphs of the graph, corresponding to identical, repeated or equivalent lines of code; see Fig.~\ref{fig:duplicate} in Appendix~\ref{sec:duplicate}.

\subsection{High-Level Description of Our Layout Algorithm}
At a high level,
there are five steps in our algorithm for producing a 
drawing of the graph, $G$:
\begin{enumerate}
\item We construct an SFR numbering and rooted
spanning tree, $T$, for our input graph, 
$G$, which will be used as the ``backbone'' of our drawing. 
\item We draw the tree $T$ using a recursive placement algorithm.

\item We add the edges of $G$ that are not in the tree $T$.

\item We highlight in our drawing 
      the sections of our graph that are most likely 
      to contain vulnerabilities, based on various criteria.

\item We automatically group subsets of nodes 
      before displaying the entire graph to the user, in a way that
      allows the user to expand such collapsed nodes. 
\end{enumerate}

In the remainder of this section, we describe in more detail each of the
above steps in our layout algorithm.

\subsection{Constructing an SFR search tree} 
\begin{figure}[ht!]
\centering
\begin{tabular}{cc}
%\parbox{0.4\textwidth}{\centering \includegraphics[width=0.32\textwidth]{if-else.pdf}\\(a)}
%	& \parbox{0.4\textwidth}{\centering \includegraphics[width=0.38\textwidth]{switch.pdf}\\(b)}
%	& \multirow{2}{*}{\parbox{0.2\textwidth}
%	{\vspace{-0.4cm}\centering \includegraphics[width=0.15\textwidth]{do-while.pdf}\\(e)}}\\

%& & \\

%\parbox{0.4\textwidth}{\centering \includegraphics[width=0.38\textwidth]{for-loop.pdf}\\(c)}
%	& \parbox{0.4\textwidth}{\centering \includegraphics[width=0.38\textwidth]{while-loop.pdf}\\(d)} & \\

\parbox{0.5\textwidth}{\centering \includegraphics[height=0.11\textheight]{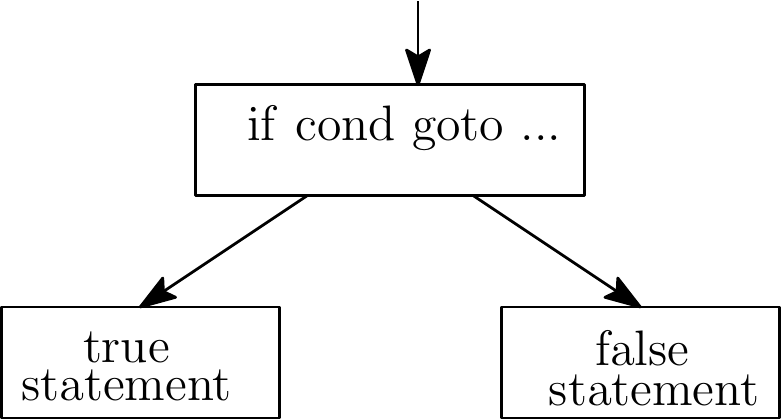}}
	& \parbox{0.5\textwidth}{\centering \includegraphics[height=0.11\textheight]{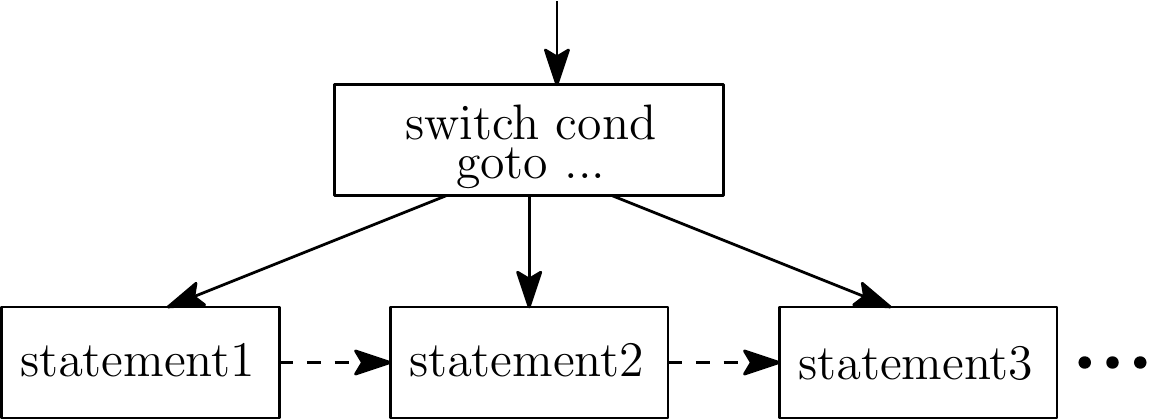}}\\

(a) & (b) \\

\parbox{0.5\textwidth}{\centering \includegraphics[height=0.11\textheight]{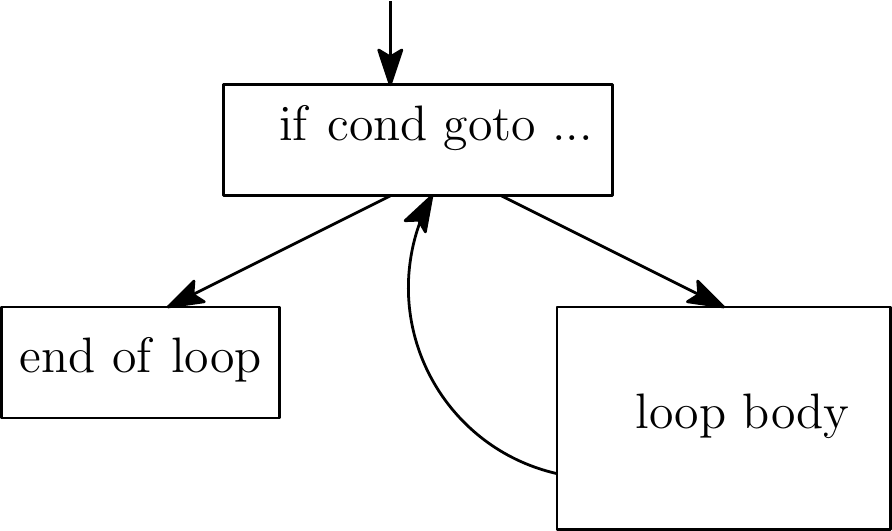}}
	& \parbox{0.5\textwidth}{\centering \includegraphics[height=0.16\textheight]{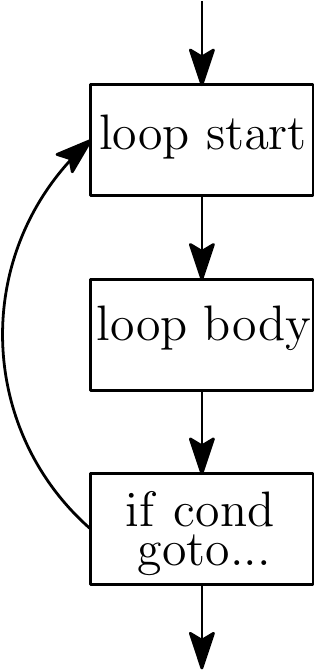}} \\

(c) & (d) \\

\end{tabular}
\caption{Sample graph layout for (a) if-else conditional statement, (b) switch statement (dashed edges for flow through statements), (c) for and while loop, (d) do-while loop.}
\label{fig:layouts}
\end{figure}

In our first step, we construct a rooted (ordered) SFR spanning tree, $T$, for
the graph, $G$, produced from the Jaam tool~\cite{jaam}.
The algorithm we use to perform this construction is exactly the
SFR algorithm shown in Fig.~\ref{fig:dfs}, 
with the added detail that
as we traverse the graph $G$ to construct our SFR search tree, $T$,
we process the out-edges from each vertex using the ordering
for $G$, consistent with the intuitive way programmers naturally organize
branches for different types of software branch points. 
As we highlight in one of our case studies, this approach tends to 
produce almost identical drawings for repeated (e.g., cut-and-pasted)
software code fragments.
It also produces ordered combinatorial layouts for each of the following
types of code constructs, as shown in Fig.~\ref{fig:layouts}.

\begin{itemize}
\item If-else conditional statement: 
the true and false components are
siblings, with the true component coming first.
\item Switch conditional statement: 
the different branches of the switch
statement are siblings of the conditional statement, ordered by their
appearance in the code (even if there are non-tree edges between them that
would be representing flow-throughs from one branch to another).
%\item For loop: 
%the increment statement and loop body are both siblings of
%the conditional statement, with the increment statement coming first and
%(because of the recusive aspect of the SFR numbering) the end-of-loop
%statement will be a child of the increment statement.
\item While/for loop:
the end-of-loop statement and loop body are both siblings of
the conditional statement, with the end-of-loop statement coming first.
\item Do-while loop: 
the start statement, loop body, and conditional
statements are in a single path, with a non-tree edge leading back to the
start statement.
% \item Exceptions
% \item Polymorphism
\end{itemize}

\subsection{Drawing the Nodes of our SFR Search Tree}
\begin{figure}[hb!]
\centering
\includegraphics[width=0.45\textwidth]{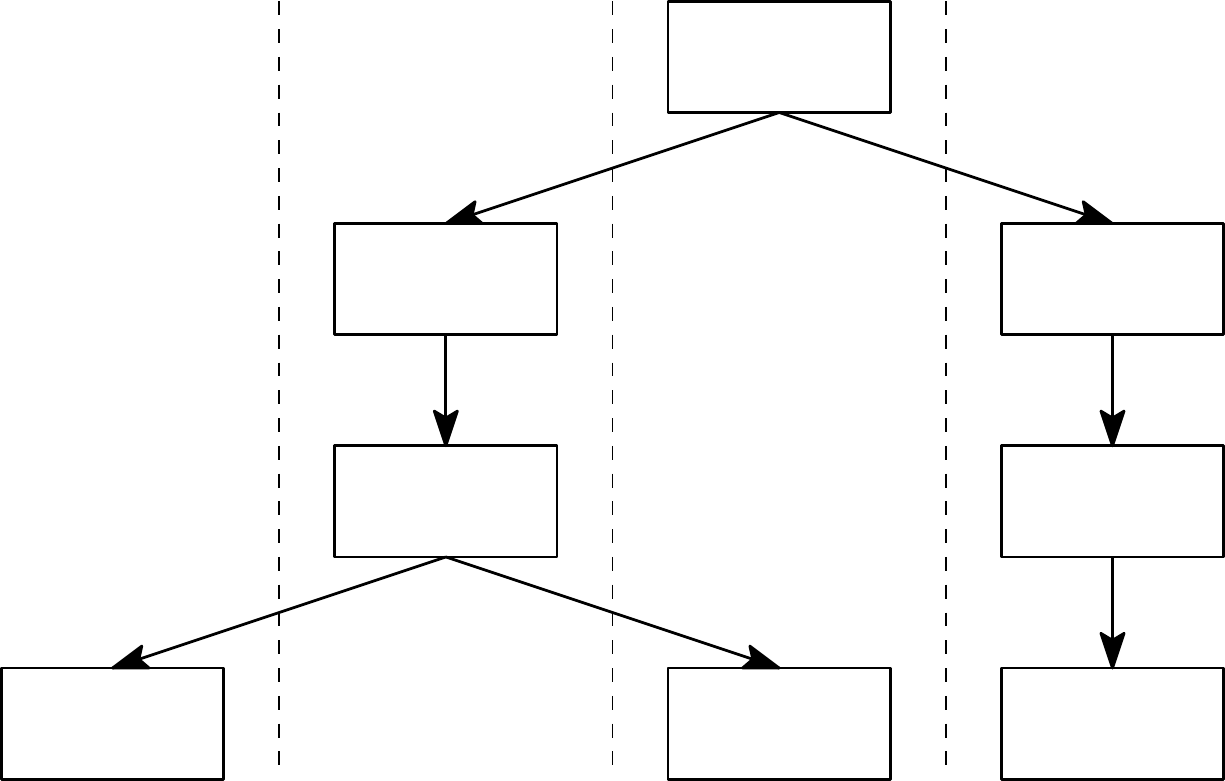}
\caption{A tree drawn according to our algorithm. The dashed vertical lines show separation of subtrees into unique lanes.}
\label{fig:tree}
\end{figure}

Once we have constructed our (ordered) SFR search tree, $T$,
we draw it recursively, starting from the leaves of $T$.  
Each parent is drawn on a row above all of its children.
Chains of nodes are drawn in a vertical column. When we reach a
branch point, we lay out each of the subtrees from left to right.
We require that only a direct descendant of a node can be placed
directly underneath it. This means that each subtree, no matter its
size, will have an entire vertical lane reserved for it from top to bottom
in our graph.
See Fig.~\ref{fig:tree}.

\begin{figure}[tb!]
\centering
\includegraphics[width=0.8\textwidth]{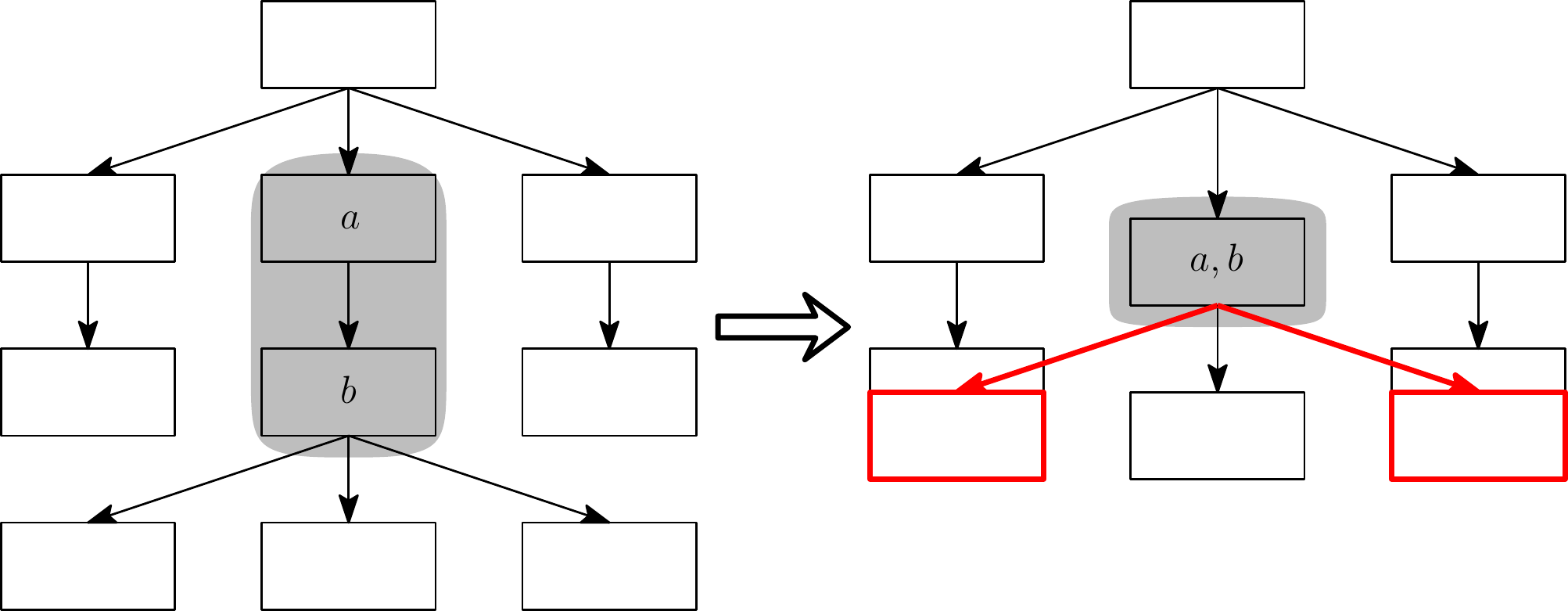}
\caption{An example of how breaking our drawing style can cause
problems. When the user collapses nodes $a$ and $b$, some of the
nodes need to be shifted. To avoid this problem, we forbid a
node from lying directly above a node that is not a direct
descendant.}
\label{fig:collapse}
\end{figure}

This requirement might at first seem to waste space in our drawing, but it 
maintains consistency when a user expands or
collapses connected sections of nodes. To see why this is so, suppose
that two nodes are placed at the same $x$-coordinate, but neither
is an ancestor of the other. 
Suppose further that
the user then chooses to collapse
the set consisting of the path from each of these nodes up to their
lowest common ancestor. 
If this happens, then if we simply shift up that portion of the spanning
tree, then we wil cause overlaps, which would require
shift of nodes to fix. 
(See Fig.~\ref{fig:collapse}.)
But such a shift would be moving nodes in a way that could be 
detrimental
to the mental map.
Thus, rather than produce a compact drawing that reuses vertical space,
we use the scheme described above, which tends to preserve the mental map
even as we would be collapsing or expanding paths in the spanning tree, $T$,
and shifting the remaining portions accordingly.

\subsection{Drawing Edges} 
After we have placed the nodes of our ordered spanning tree, $T$,
we must draw all of the edges in our graph. 
In our case
we choose to draw downward edges as straight line segments, 
and we then draw upward edges as curved segments.
In addition to drawing arrows at the ends of such segments,
this convention provides a visual cue for which direction an edge
is pointing. 
It also prevents upward edges from lying on top of
downward edges. For example, this makes the drawing of
the graph of software implementing the bubblesort 
algorithm, shown in Fig.~\ref{fig:bubblesort} in Appendix~\ref{sec:examples},
more readable.

\subsection{Highlighting} 
After the nodes and edges of the graph, $G$, are placed,
we highlight vertices to guide the user on where to begin
examining it to discover possible security vulnerabilities. 
For attacks based on increasing the running time for the software
on certain inputs, we want to highlight nodes that are
likely to be visited the greatest number of times during an execution.
So we color our states from green to red based on how
likely each node is to be involved in a vulnerability. 
There are several ways to determine such a vulnerability 
ranking. 
We have considered the following three:
\begin{itemize}
\item 
We could perform a recurrence analysis that computes an upper bound
on how many times each node is visited. Automated recurrence analysis
has progressed to the point of being able to provide strong upper
bounds on many simple algorithms~\cite{albert2007costa}. But we do
not believe that this is yet possible for programs such as the ones
we need to examine, which contain thousands or tens of thousands
of lines of code, and have a complex loop structure. 
\item 
We could profile our code, by providing a sample input and counting
how many times each state is visited. But while this may help for
honestly written programs, we believe it is unlikely to identify
the kinds of deliberate attacks that we need to discover. The
programs we are given should in most cases perform well, because
otherwise they would not be used at all. But some may have a hidden
trigger that causes them to run for much longer.

\item 
We could count the number of nested loops that contain each node.
This is a somewhat naive method, but it does seem to give a reasonable
heuristic, as will be seen in our case studies. It is also feasible
to compute even for very large graphs. Hence this is the method
that we choose to use.
\end{itemize}

In order to do this highlighting, of course, we need to determine for
each node its level of nesting with respect to the loops of the program.
We use 
an adaptation to the SFR tree of a definition 
by Havlak~\cite{havlak1997nesting} for DFS trees:
\begin{itemize}
\item 
The outermost loops are the maximal strongly connected components of
the directed control-flow graph, $G$.
\item 
The header for a loop is the first node in the loop that is reached in 
the SFR tree, $T$.
\item 
The inner loops are the maximal strongly connected components 
that remain when the header is removed.
\end{itemize}

A graph is said to be \emph{reducible} 
if every cycle has a single entry point~\cite{havlak1997nesting}.
If the graph is reducible, then the loop decomposition does not
depend on which rooted spanning is used. But if a cycle has
multiple entry points, then the order in which we explore the
branches could matter. Thus, in our case,
we use the canonical SFR tree, $T$, 
that has already been defined for our graph.

To compute loop headers efficiently, we use an algorithm from
Tao {\it et al.}~\cite{wei2007new}. This traverses the ordered spanning
tree and passes loop header information up the tree. While their
method could take a long time for artificially complex graphs, it
takes linear time for most real-world programs, because the spanning
trees for such graphs tend to be reducable or ``nearly'' reducible.

\subsection{Grouping Nodes}
We have included four different ways in which nodes in the graph, $G$
can be aggregated, and we present the initial drawing of 
$G$ to the user based on a pre-defined grouping 
of certain nodes, with some of these pre-collapsed.
First, we choose not to explore nodes that correspond to calls to the
Java library, since we do not expect it to contain vulnerabilities.
Instead, every such call is collapsed to a single line. This prevents
us from creating hundreds of thousands of nodes for the Java library.
Nevertheless,
the static analyzer, Jaam,
needs to do this carefully, so that it can approximate the state
of Java library objects and predict their later behavior. For
example, any object that is added to an ArrayList or a HashMap can
later be taken out.
Still, we assume that such an identification is given as an annotation
to the input graph, $G$, since this identification is solely the domain
of the Jaam tool.
Second, we automatically group each connected set of nodes that
belong to the same method. That way, if the user is not interested
in the details of a given method, they can collapse it to a single
node.
Third, we automatically group chains of method nodes that were
created in the previous step. Generally, having long chains of nodes
taking up a large portion of the screen space hinders the user from
seeing the branching structure that they need to find.
Finally, we allow the user to select a connected set of nodes and
collapse them dynamically, 
along with providing a comment explaining the purpose of 
the corresponding section of code.

\section{The User Interface}
In constructing the user interface for the J-Viz system, we relied in
part on a survey study of Basil and Keller~\cite{bassi2001software}
for software visualizations.
They found that the following criteria were considered ``absolutely
essential'' by a majority of participants:
\begin{quotation}
\noindent
$\bullet$ Search tools for graphical and/or textual elements
\hfil\break $\bullet$ 
Source code visualization
\hfil\break $\bullet$ 
Hierarchical representations
\hfil\break $\bullet$ 
Use of colors
\hfil\break $\bullet$ 
Source code browsing
\hfil\break $\bullet$ 
Navigation across hierarchies
\hfil\break $\bullet$ 
Easy access from the symbol list of the corresponding source code.
\end{quotation}

To make our system more user-friendly, we have implemented each of
these features. We have already discussed our hierarchy for collapsing
and expanding vertices, and our use of colors for highlighting
nested loops in our graph. We show the other features in use in
the screenshot shown in
Fig.~\ref{fig:full-system} in Appendix~\ref{sec:full-system}, and describe each of them below.

\subsection{Searching} 
We have included the following searching capabilities in our graph:
\begin{quotation}
\noindent
$\bullet$ Find nodes by SFR numbering
\hfil\break$\bullet$ Find incoming and outgoing edges for a given node
\hfil\break$\bullet$ Find nodes whose methods contain a given string
\hfil\break$\bullet$ Find nodes whose instructions contain a given string
\end{quotation}

\subsection{Context panel} 
When nodes are selected by the user, the code for their methods are shown in the left panel, with the lines for each node highlighted. Alternatively, a user can select lines from the left panel, and we highlight the nodes in our graph that correspond to those lines of code. The lines of code for the innermost loops inside each method are also highlighted.

\subsection{Description panel}
The right panel displays detailed descriptions for each of the selected nodes in our graph.

\subsection{Minimap} 
A minimap is given in the lower left that always shows our full graph. When we zoom in on part of the graph, it is highlighted on the map, so that the user never loses track of where they are.

\section{Case Studies}

In this section, we describe some case studies we performed to test 
the effectiveness of the J-Viz system for visualizing Java bytecode
and identifying security vulnerabilities that could be triggered
by algorithmic complexity attacks.
As input to these case studies,
we were provided by DARPA with programs to analyze to test our system,
some of which were produced by a ``red team'' tasked with deliberately
creating software that contains vulnerabilities to 
algorithmic complexity attacks. 

Our first case study is for a program for verifying a password,
which is meant to be kept secret, and not revealing any information
about such a password, including whether
it is valid or not.
In this case, J-Viz was effective in leading a security analyst 
to a nested loop that checks each character in the password one at a time. 
In part, because of the way that the SFR spanning tree lays out conditional 
branches in an intuitive manner and draws
loop edges as curved segments, the analyst was able to notice
that the password-checking program exits as soon as it finds the
first character that does not match.
(See Fig.~\ref{fig:pwcheck} in Appendix~\ref{sec:full-system}.)
The analyst then correctly identified this as a vulnerability (inserted
by the red team), since,
by timing multiple executions of the program, an attacker can easily
determine how many correct characters of a password
that they entered with each attempt.
Thus, such an attacker could quickly crack the password by a simple
iterative search.

\begin{figure}[hbt]
\begin{center}
\includegraphics[width=4in, height=3in]{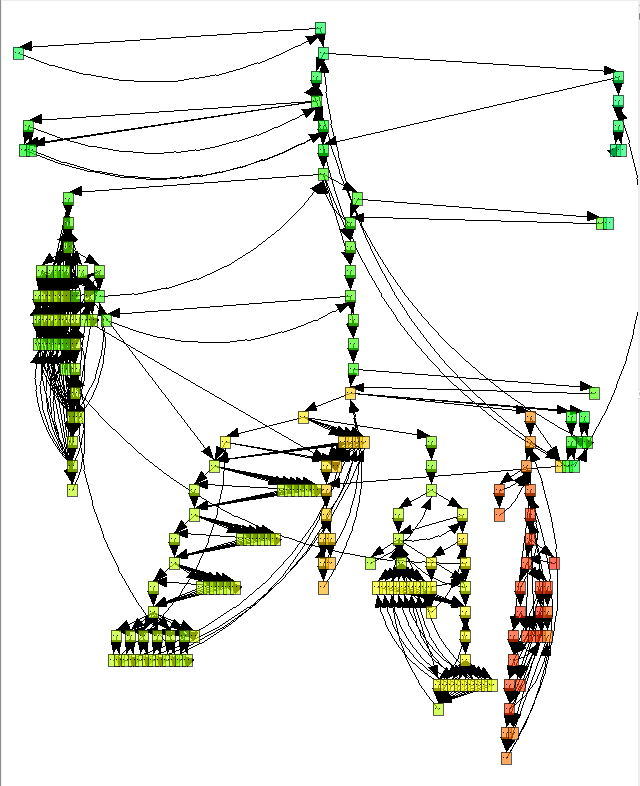}
\end{center}
\caption{Our visualization for the graph for an image processor program, 
showing the red section of a deeply nested loop
(which contained a security vulnerability) in the bottom right.}
\label{fig:image-processor}
\end{figure}

Our second case study is for a program for analyzing and classifying images
based on features, such as the number of edges or the amount of each
color that is present. This program is around 1,000 lines of Java
code, and produces about 3,000 nodes in our graph. 
The goal of the security analyst in this case was to
determine if this program can be made to take much longer than it should
(specifically, greater than 18 minutes to analyze a 70 KB image).
For most images of that size, the program takes around 6 minutes.
But, through the use of J-Viz, a security analyst
was able to create an image that would take over an hour to be
analyzed.
The key insight for the analyst
was to pay attention to the highlighting in our visualizer,
which showed the deepest nested loops in dark red.
(See Fig.~\ref{fig:image-processor}.) 

\begin{figure}[hbt]
\begin{center}
\includegraphics[scale=0.4]{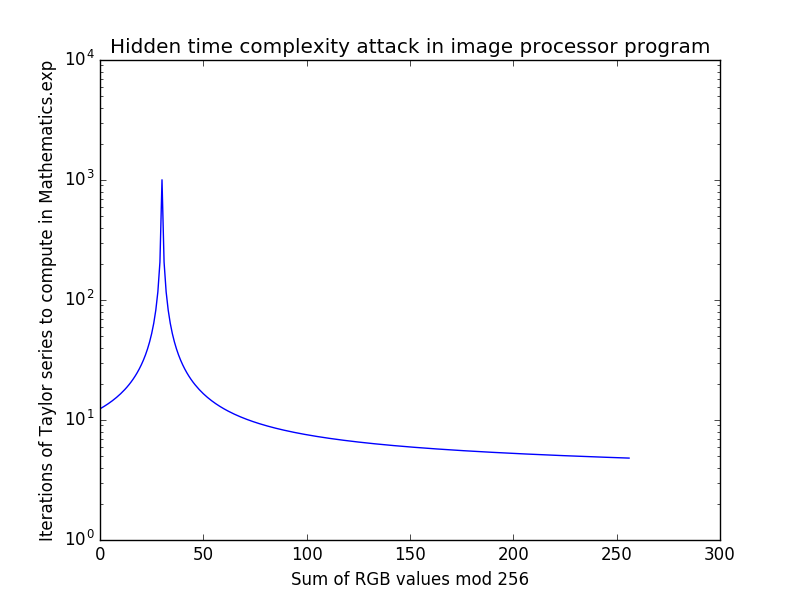}
\end{center}
\caption{The time complexity of image processor spikes by two orders of magnitude when the sum of the RGB values is exactly 30.}
\label{fig:image-processor-accuracy}
\end{figure}

In this case, the analyst
noticed that many of these
nodes were part of the Mathematics class, which provided custom
implementations of various mathematical functions.
In particular, the exponential function was implemented using a
Taylor series, with the number of terms depending on a function of
the RGB value of each pixel. 
This function was highly non-linear and contained a ``spike,''
which is shown in Figure~\ref{fig:image-processor-accuracy},
which implied a large number of terms in the Taylor series for an
image having a particular RGB value.
Given this information,
the analyst was then able to create an image
that triggered this behavior for every pixel, which took over an
hour to process, confirming the vulnerability.

In addition to these case studies, we show additional examples and screenshots
in Appendix~\ref{sec:examples}~and~\ref{sec:full-system}.

\section{Conclusion and Future work}
We have described a new software visualization tool, J-Viz, 
that uses an SFR numbering scheme for graphs produced using the 1-CFA framework
so that similar sections of code should be drawn similarly and deeply
nested code portions are placed well and highlighted. 

In keeping with recent graph drawing research, we have argued that
our system preserves the mental map of the user as they interact
with the graph. We also meet all of the
criteria that both programmers and researchers have considered
essential for software visualization. As a result, our system has
already proven to be useful for human analysts in finding various kinds of
software vulnerabilities.

In future work, we plan to 
test our system for larger programs. 
We also plan to study ways to provide semi-automated methods for 
identifying other kinds of potential algorithmic-complexity
security vulnerabilities
rather than simply counting the number of nested loops. 

\subsection{Acknowledgements}
This article reports on work supported by the Defense Advanced
Research Projects Agency under agreement no.~AFRL FA8750-15-2-0092.
The views expressed are those of the authors and do not reflect the
official policy or position of the Department of Defense or the
U.S.~Government.
This work was also supported in part by the U.S.~National Science Foundation
under grants 1228639 and 1526631.
In addition, we would like to thank David Eppstein, Matthew Might, 
William Byrd, Michael Adams, and Guannan Wei for helpful discussions regarding the topics of this paper.

%\bibliographystyle{splncs03}
%\bibliography{refs}

\input{main.bbl}
\clearpage
\input{appendix}

\end{document}

%% file: appendix.tex
\begin{appendix}

\section{SFR Numbering and Isomorphic Subgraphs}
\label{sec:duplicate}

\begin{figure}[htb]
\centering
\includegraphics[width=0.8\textwidth]{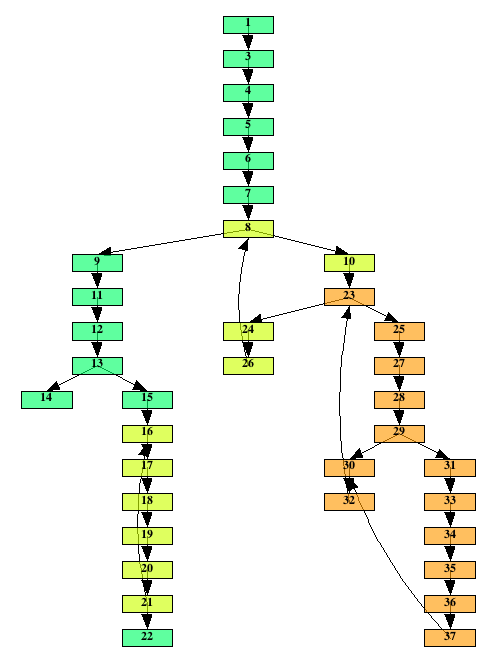}
\caption{A graph for the bubblesort algorithm; vertices are labeled with SFR numbers.}
\label{fig:sfr-number}
\end{figure}

Fig.~\ref{fig:sfr-number} illustrates a graph (for the complete code of a bubblesort implementation) where the vertices are labeled with their SFR numbers.

\newpage

\begin{figure}[htb]
\centering
\includegraphics[width=0.6\textwidth]{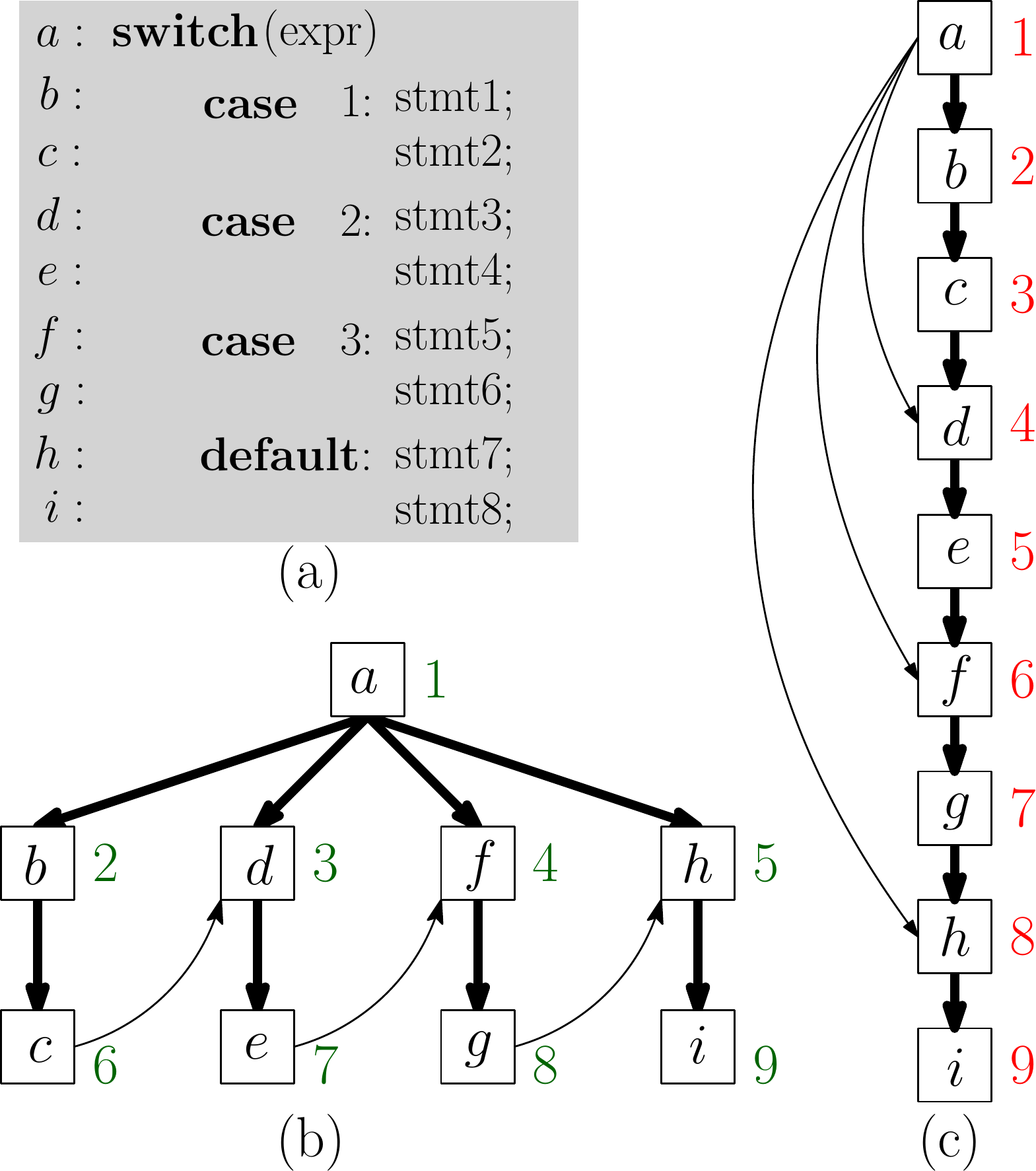}
\caption{(a) A code segment containing a switch statement, (b) the spanning tree (thick edges) and
 vertex numbering (in blue color) for the graph obtained from SFR search, and (c) the spanning tree
 (thick edges) and vertex numbering (in red color) for the graph obtained from depth-first search.}
\label{fig:sfr}
\end{figure}

Fig.~\ref{fig:sfr} illustrates the difference in the spanning trees and the vertex numberings obtained
 in our SFR search, and a more conventional depth-first search. Both the trees and vertex numbering
 is for a code segment containing switch statement. The tree and the numbering obtained in the SFR
 shows the structure of the switch statement more naturally.

\newpage

\begin{figure}[htb]
\centering
\includegraphics[width=\textwidth]{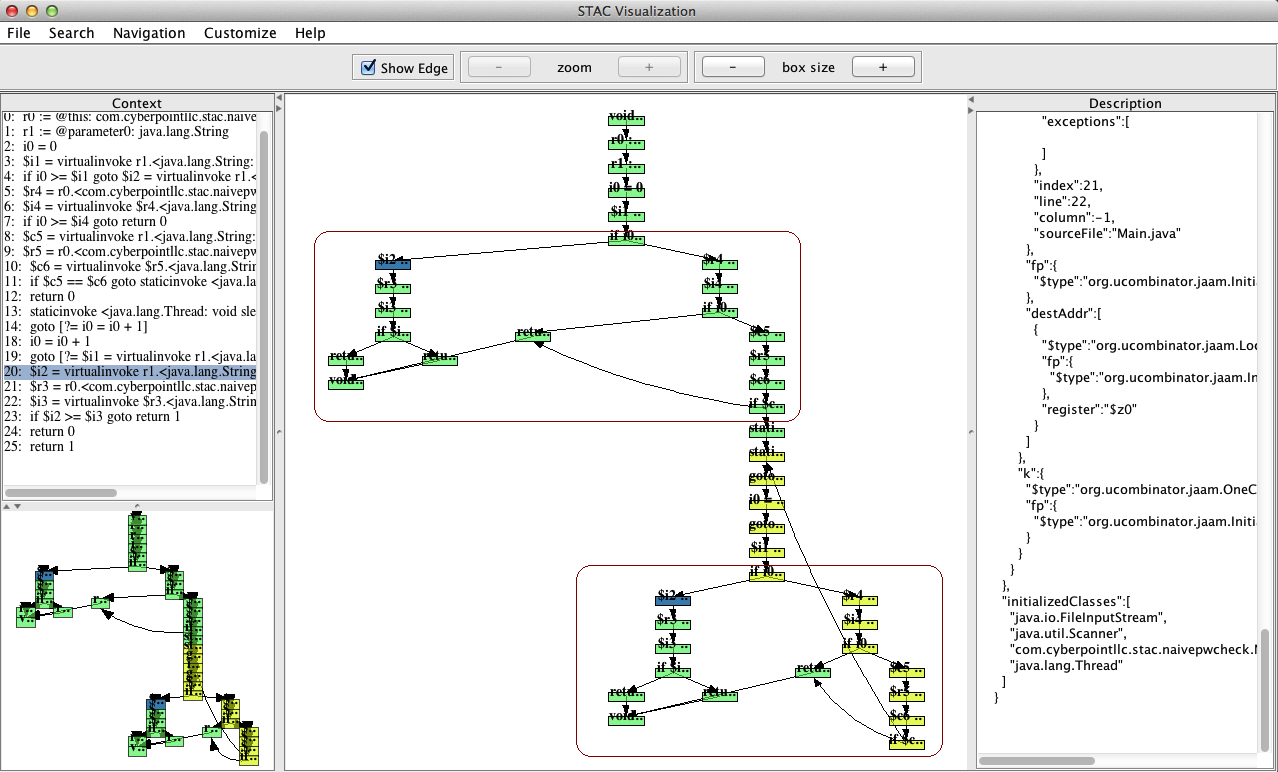}
\caption{Visually separable isomorphic subgraphs in the graph for the password checker code; these isomorphic subgraphs correspond to the same or equivalent lines of code.}
\label{fig:duplicate}
\end{figure}

The SFR search tree and the SFR numbering also enables viewers to visually identify isomorphic
 subgraphs in the graph, which correspond to identical or equivalent lines of code in the program;
 see Fig.~\ref{fig:duplicate}.

\newpage

\section{Additional Figures and Examples}
\label{sec:examples}

In this we show the graphs for some simple algorithms, as visualized in our J-viz. In particular we show the graphs for a bubblesort algorithm, for two simple programs implementing the recursive functions computing the factorial of a number and the $n$-th Fibonacci number, respectively.

\medskip

\noindent
\textbf{Bubblesort Algorithm}

\begin{figure}[h]
\centering
\includegraphics[scale=0.7]{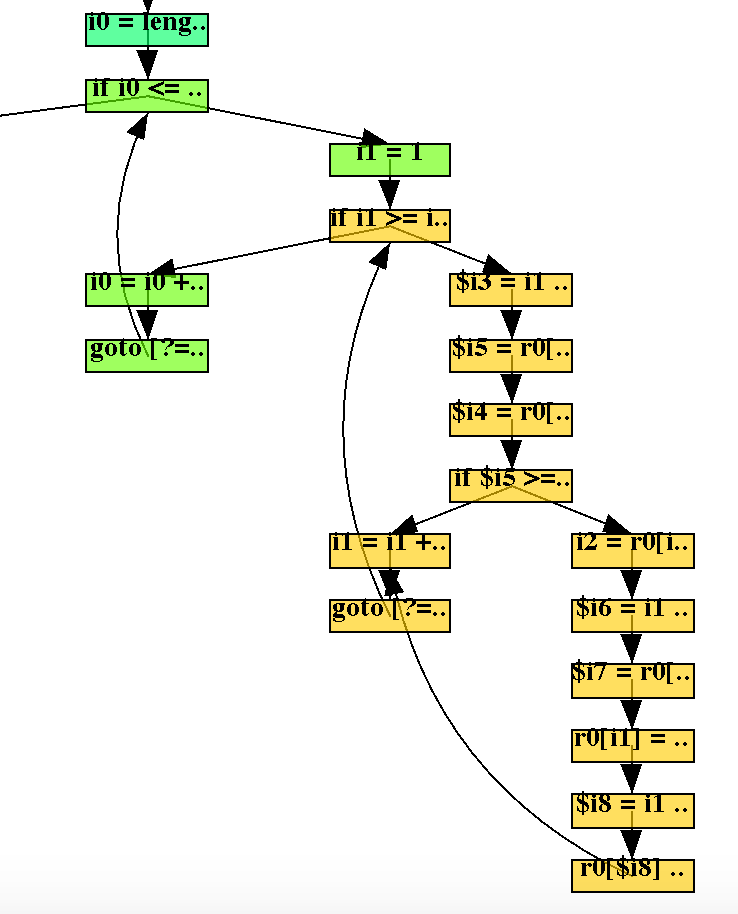}
\caption{This partial graph of a bubblesort algorithm shows how drawing upward edges as curves and highlighting nested loops can improve readability. It also shows our use of colors for different levels of nested loops.}
\label{fig:bubblesort}
\end{figure}

\newpage
\noindent
\textbf{Computation of Factorial($n$)}

\begin{figure}[htb]
\centering
\includegraphics[scale=0.5]{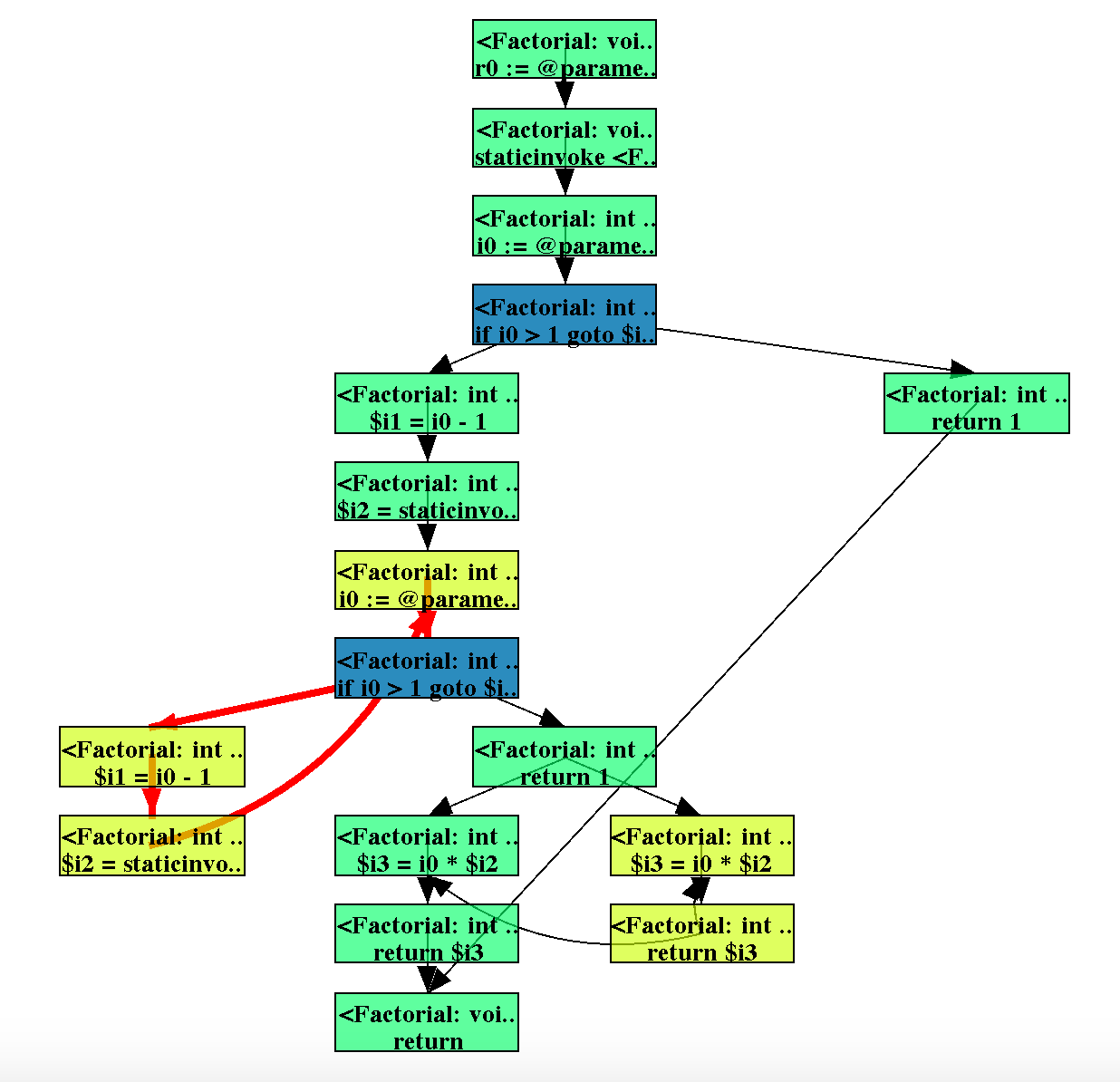}
\caption{A graph of a recursive factorial function using 1-CFA. The highlighted instruction occurs twice - once in the context of being called from main, and once in the context of the factorial function calling itself.
Note, in addition, how this tree is different from a breadth-first search
(BFS) tree. Namely, there is a long forward edge from the rightmost node
in the drawing. In a BFS numbering of this graph, that edge would force
its end-vertex to a higher level, which would distort the natural notion
of recursive depth that the SFR spanning tree illustrates better here.
}
\label{fig:factorial-drawing}
\end{figure}

\newpage
\noindent
\textbf{Computation of the $n$-th Fibonacci Number}

\begin{figure}[htb]
\centering
\includegraphics[scale=0.5]{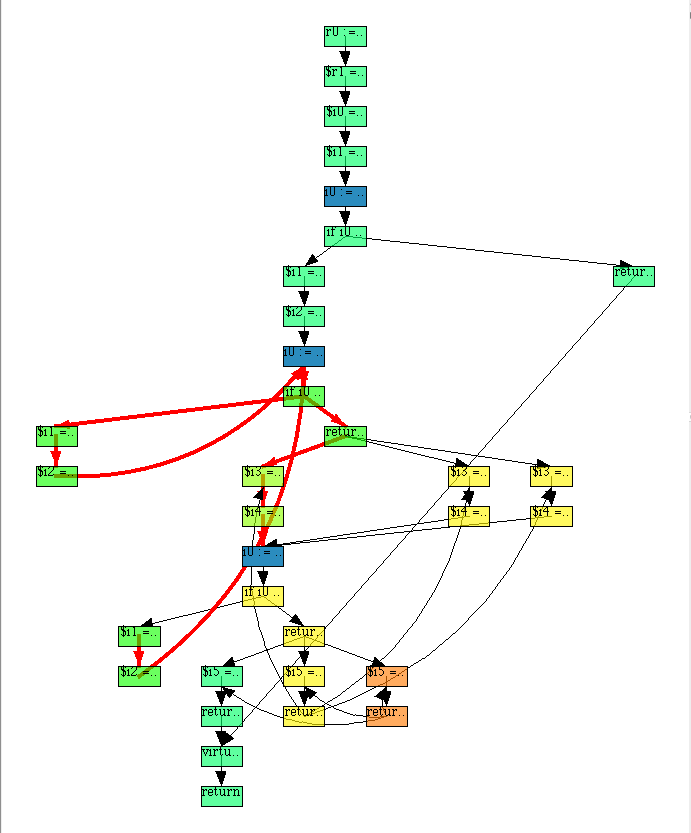}
\caption{A graph of a recursive Fibonacci function using 1-CFA. The highlighted instruction occurs three times, because the function can be called from main, or from itself in two different places.}
\label{fig:Fib-drawing}
\vspace{-2cm}
\end{figure}

\newpage

\section{Illustration of the J-Viz System}
\label{sec:full-system}

\begin{figure}[hbt]
\centering
\includegraphics[scale=0.3]{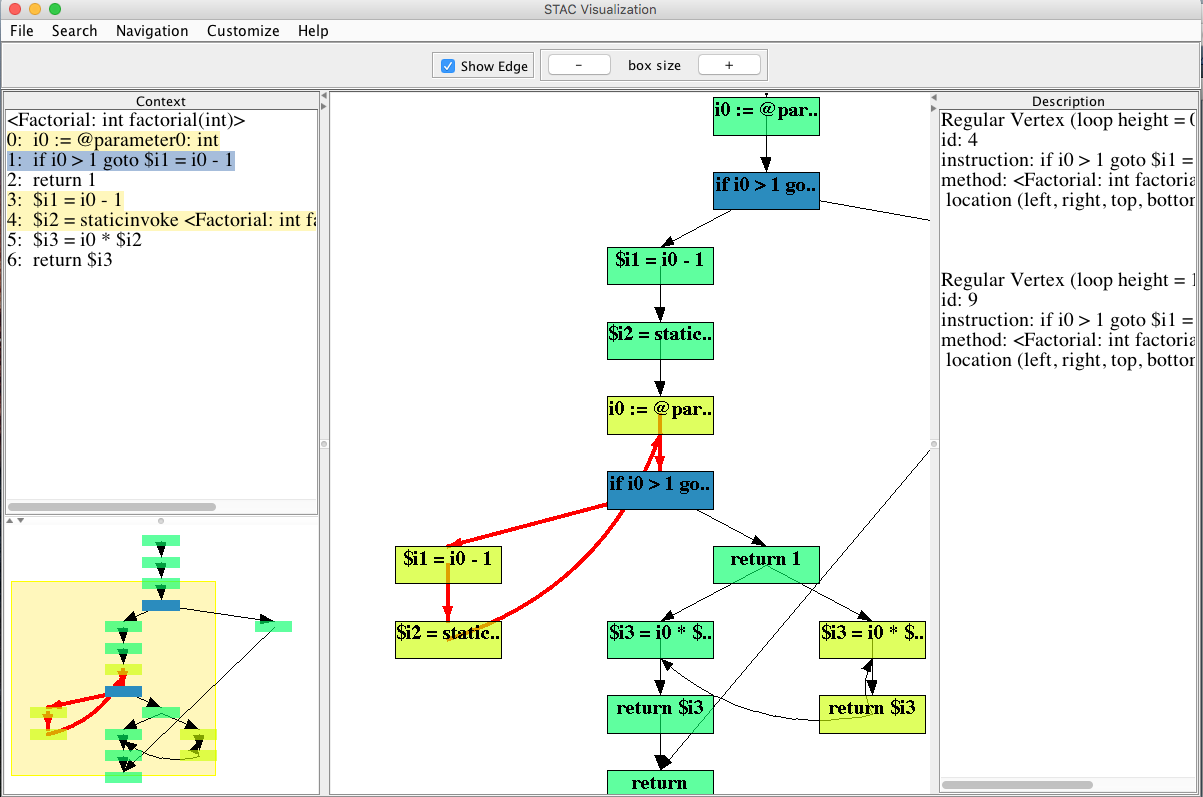}
\caption{A screenshot of our complete visualization system, J-Viz,
with a (clipped) view of an example Factorial program. This shows the left panel with the code for the currently selected method, the right area with the description of each selected node, and the minimap with our current zoom level.}
\label{fig:full-system}
\end{figure}

Fig.~\ref{fig:full-system} illustrates our J-viz system and its different component.

\newpage

\begin{figure}[htb]
\centering
\includegraphics[width=\textwidth]{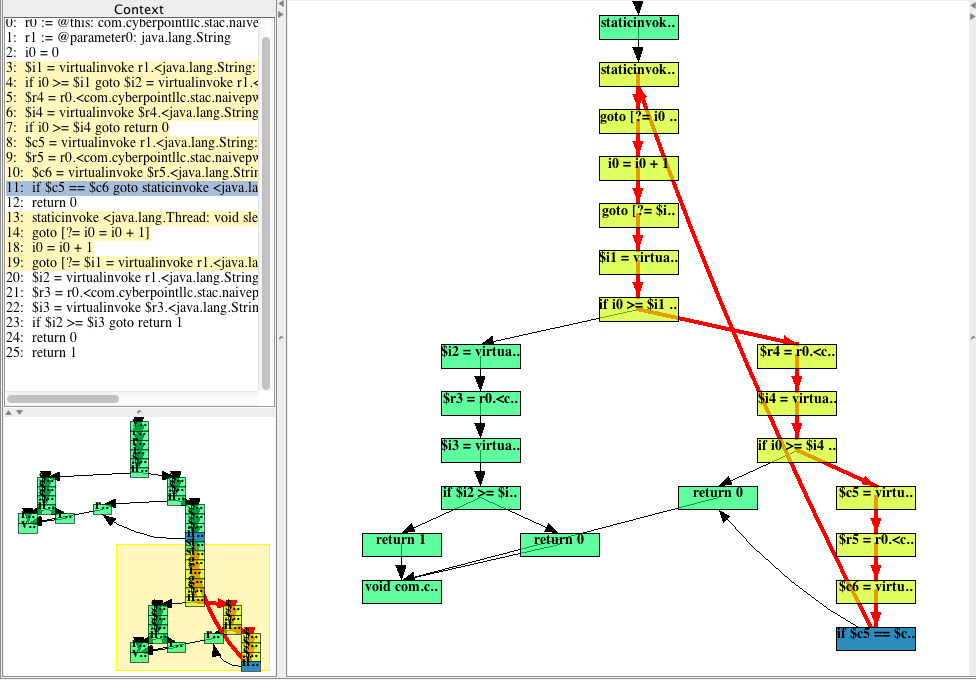}
\caption{Our layout of the graph for a password checker with the relevant part zoomed, as a part
of our first case study. 
The highlighted node shows a check for each character of a password. 
If this fails, the program exits the loop immediately, 
allowing for a side-channel attack (for identifying failed passwords).}
\label{fig:pwcheck}
\end{figure}

Fig.~\ref{fig:pwcheck} illustrates how the J-viz system is used in indentifying vulnerable code segments in a password checker program.

\end{appendix}